\documentclass[prl,twocolumn,showpacs,superscriptaddress,preprintnumbers]{revtex4}
\usepackage{amssymb}
\usepackage{graphicx}
\usepackage{dcolumn}
\usepackage{bm}
\usepackage[bookmarks = true, pdfpagemode = None, pdfstartview = FitH,colorlinks = true,
citecolor = black, linkcolor = black, urlcolor = black]{hyperref}
\usepackage{url}

\begin{document}

\title{Probing Noise in Flux Qubits via Macroscopic Resonant Tunneling}
\author{R.~Harris}
\email{rharris@dwavesys.com}
\author{M.W.~Johnson}
\affiliation{D-Wave Systems Inc., 100-4401 Still Creek Dr., Burnaby, BC V5C 6G9, Canada}
\homepage{www.dwavesys.com}
\author{S.~Han}
\affiliation{Department of Physics and Astronomy, University of Kansas, Lawrence KS, USA}
\author{A.J.~Berkley}
\author{J.~Johansson}
\author{P.~Bunyk}
\author{E.~Ladizinsky}
\author{S.~Govorkov}
\author{M.C.~Thom}
\author{S.~Uchaikin}
\affiliation{D-Wave Systems Inc., 100-4401 Still Creek Dr., Burnaby, BC V5C 6G9, Canada}
\author{B.~Bumble}
\author{A.~Fung}
\author{A.~Kaul}
\author{A.~Kleinsasser}
\affiliation{Jet Propulsion Laboratory, California Institute of Technology, Pasadena CA, USA}
\author{M.H.S.~Amin}
\affiliation{D-Wave Systems Inc., 100-4401 Still Creek Dr., Burnaby, BC V5C 6G9, Canada}
\author{D.V.~Averin}
\affiliation{Department of Physics and Astronomy, SUNY Stony Brook, Stony Brook NY, USA}
\date{\today }

\begin{abstract}
Macroscopic resonant tunneling between the two lowest lying states of a
bistable RF-SQUID is used to characterize noise in a flux qubit.  Measurements of the incoherent decay rate as a function of flux bias revealed a Gaussian shaped profile that is not peaked at the resonance point, but is shifted to a bias at which the initial well is higher than the target well.  The r.m.s. amplitude of the noise, which is proportional to the decoherence rate $1/T_2^*$, was observed to be weakly dependent on temperature below $70\,$mK.  Analysis of these results indicates that the dominant
source of low frequency ($1/f$) flux noise in this device is a quantum mechanical environment in thermal equilibrium.
\end{abstract}

\maketitle


The viability of any scalable quantum computing architecture is
highly dependent upon its performance in the presence of noise. In
the case of superconducting qubits it has been shown that low
frequency (1/$f$) flux noise is of particular concern \cite{1overF}.
Furthermore, evidence suggests that such devices  generically couple
to an ensemble of effective 2-level systems (TLS) that may be
materials defects \cite{Martinis1}. A number of theories exist that
attempt to correlate these two observations \cite{TLStheory},
however it is not certain whether the TLS observed in spectroscopy
experiments are \textit{the} dominant source of low frequency noise
in these devices \cite{Martinis2}. The development of additional
experimental probes of low frequency noise will prove critical in
the quest to build reliable superconductor-based quantum computing
hardware.  In this article we demonstrate a new experimental
procedure for quantifying low frequency flux noise in RF-SQUID
qubits.  The procedure developed herein complements other approaches
to using qubits as spectrometers for studying noise \cite{Schoelkopf}.

Macroscopic resonant tunneling (MRT) \cite{Averin00,MRT} is an
important probe of quantum effects in Josephson junction-based
devices. In an MRT experiment flux tunnels between two wells of a
double well potential when energy levels are aligned. The tunneling rate
and width of the tunneling region are strongly influenced by the environment.
The effect of flux noise on the two lowest energy levels can be described using
an effective Hamiltonian
\begin{equation}
\label{eqn:Ham}
{\cal H}_{\text{eff}}=-{\frac{1}{2}}\left[\epsilon\sigma_z+\Delta\sigma_x\right]-\frac{1}{2}Q_z\sigma_z\; ,
\end{equation}

\noindent where $\epsilon$ is the bias energy between the wells, $\Delta$ is the tunneling amplitude, $Q_{z}$ is an operator that acts on the low frequency modes of the environment and $\sigma_{x(z)}$ are Pauli matrices.  In general, a transverse coupling to the environment should also exist but it is believed to be subdominant to the longitudinal coupling in flux qubits \cite{Noise}.  It is demonstrated that the experimental results presented herein are consistent with this expectation.

A theoretical analysis of MRT in the presence of low frequency (non-Markovian) flux noise was reported in Ref.~[\onlinecite{Amin1}].  The transition rate $\Gamma_{01}$ from state $\left|0\right>$ to state $\left|1\right>$ (both eigenfunctions of $\sigma_z$ with eigenvalues $\mp1$, respectively) was found to be
\begin{equation}
\label{eqn:Amin1}
\Gamma_{01}(\epsilon)=\sqrt{\frac{\pi}{8}}\frac{\Delta^2}{W} \exp\left[-%
\frac{(\epsilon-\epsilon_p)^2}{2W^2}\right]\; ,
\end{equation}
\vspace{-10pt}
\begin{displaymath}
\epsilon_p={\cal P}\int_{-\infty}^{\infty} d\omega\frac{S(\omega)}{%
\omega}\; , \quad W=\left[\int_{-\infty}^{\infty} d\omega S(\omega)
\right]^{1/2} \; ,
\end{displaymath}

\noindent where $S(\omega)=\frac{1}{2\pi}\int dt\ e^{i\omega t}\langle
Q_z(t)Q_z(0)\rangle$ is the unsymmetrized noise spectral density \cite{weiss}.
$\Gamma_{10}$ is obtained by substituting $\epsilon_p\to -\epsilon_p$.
Equation (\ref{eqn:Amin1}) is valid provided the noise spectrum is peaked at low frequency and the integrals defining $\epsilon_p$ and $W$ are finite \cite{Amin1}. The latter constraint may require low and high frequency cutoffs.

The quantities $\epsilon_p$ and $W$ represent the energy shift and width of
a Gaussian shaped tunneling rate, respectively. $W$ is the
r.m.s.~amplitude of the noise and $\epsilon_p$ is a measure of the
asymmetry of $S(\omega)$. For a classical noise source $S(\omega){\approx}%
S(-\omega)$, hence $\epsilon_p = 0$.  For a quantum source $S(\omega)$ need not be symmetric and therefore $\epsilon_p\neq 0$.  In thermal equilibrium at temperature $T_{\text{eff}}$ the fluctuation-dissipation theorem dictates that $S(\omega)$ can be expressed as a sum of symmetric and antisymmetric components $S(\omega)=S_s(\omega)+S_a(\omega)$, where $S_s(\omega)=S_a(\omega)\coth\left(\omega/2T_{\text{eff}}\right)$ ($\hbar/k_B\equiv 1$).  If $S_a(\omega)$ is sharply peaked near $\omega=0$ then it can be shown that
\begin{equation}
\label{Teff}
W^2\approx{\cal P}\int_{-\infty}^{\infty}d\omega S_a(\omega)(2T_{\text{eff}}/\omega)=2T_{\text{eff}}\epsilon_p \; .
\end{equation}

The width $W$ is an important parameter in adiabatic quantum
computation \cite{AQC} as it defines the precision to which a target
Hamiltonian can be specified in the logical basis defined by the eigenstates of
$\sigma_z$. From the perspective of gate model quantum computation
$W$ is closely related to the decoherence time $T_2^*$ in the energy basis defined by the eigenstates of Eq.~(\ref{eqn:Ham}). For a low frequency
environment the decay due to dephasing has the form
$e^{-t^2/2T_2^{*2}}$, where $1/T_2^{*}=\cos\left(\eta\right) W$ and $\eta = \arctan (\Delta/\epsilon) $.


To demonstrate the MRT method for characterizing low frequency noise
we employed an RF-SQUID qubit \cite{Leggett,RFSQUIDQUBIT}. Previous
experimental observations of MRT in RF-SQUIDs have been limited to
tunneling into higher energy levels with and without the help of microwave activation \textrm{\cite{MRT,Friedman00}}.  In this paper we examine MRT
between the two lowest energy levels of an RF-SQUID.  Measurements of directional tunneling rates $\Gamma _{01}$ and $\Gamma _{10}$ are used to extract quantitative information regarding $S(\omega)$.

\begin{figure}[tbp]
\includegraphics[width=3.3in]{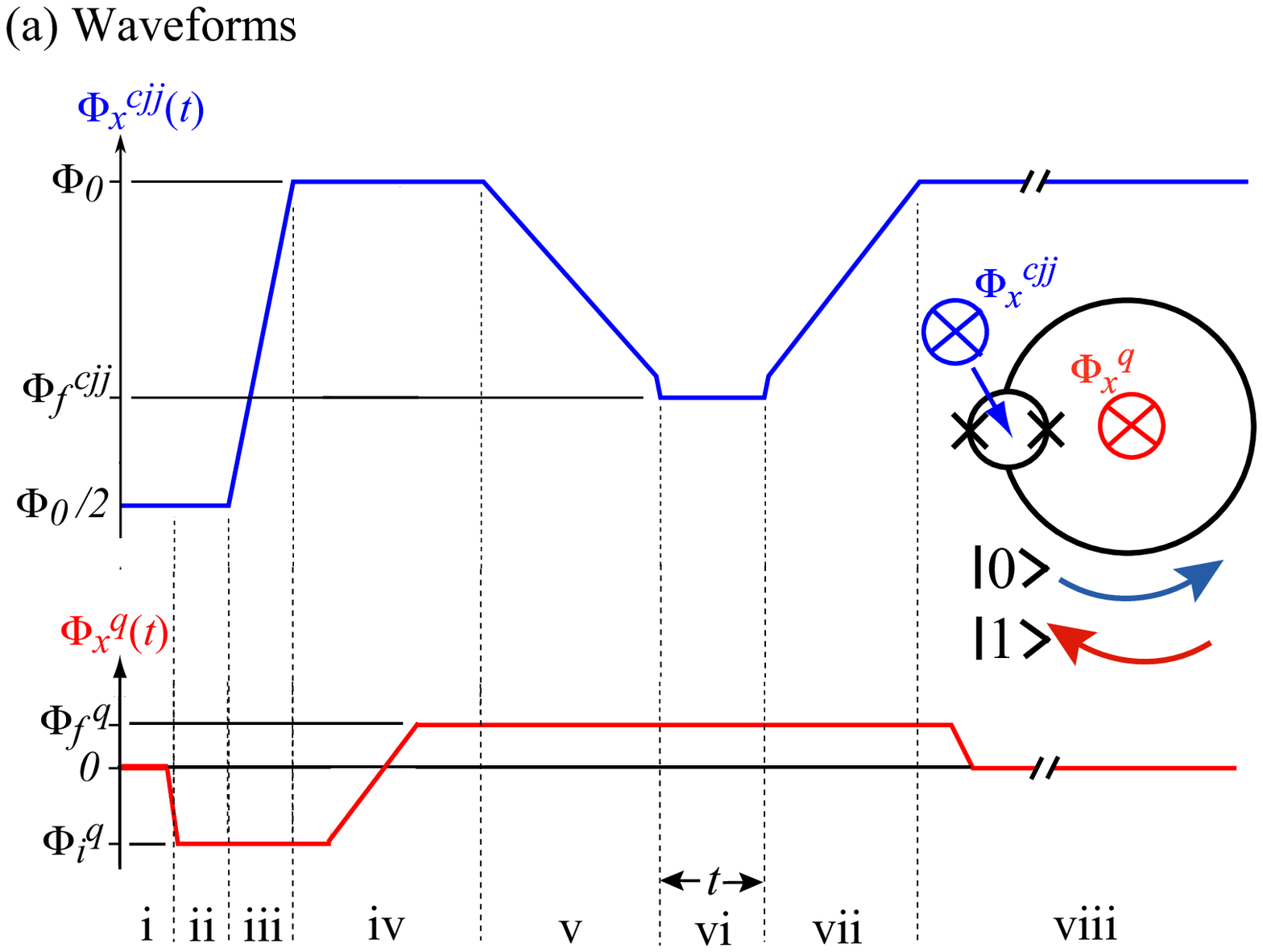}\\
\includegraphics[width=3.3in]{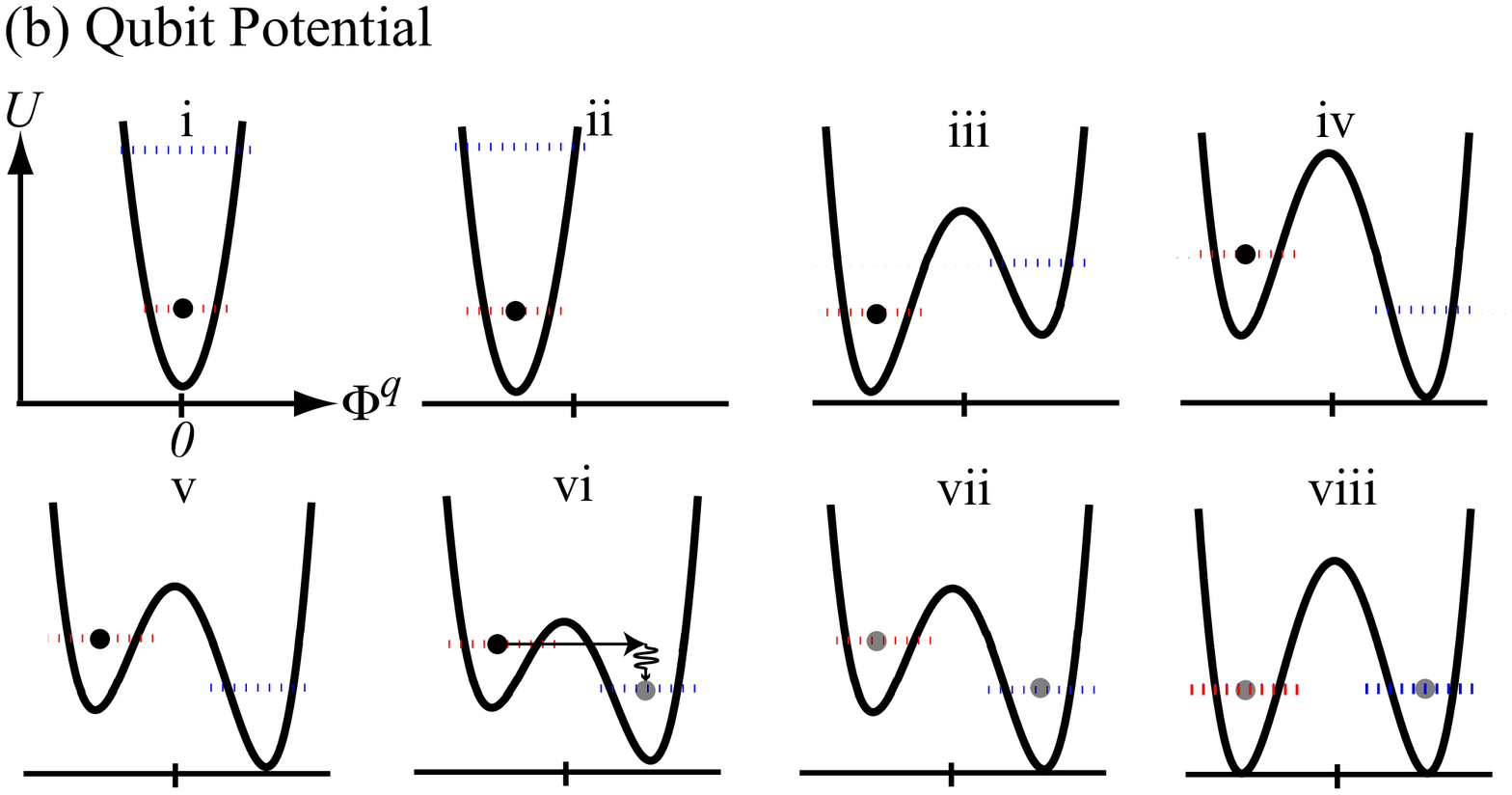}
\caption{(Color online) (a) MRT flux control bias sequence as a function of time.  Inset depicts a CJJ RF-SQUID subjected to the external biases $\Phi_x^{\text{\textit{cjj}}}$ and $\Phi_x^{\text{\textit{q}}}$.  Sense of the macroscopic persistent current states $\left|0\right>$ (counterclockwise) and $\left|1\right>$ (clockwise) are noted. (b) Evolution of the qubit potential $U(\Phi^q)$.}
\label{fig:ControlSequence}
\end{figure}

A schematic of a compound Josephson junction (CJJ) RF-SQUID is shown in the inset of Fig.~\ref{fig:ControlSequence}(a).  It consists of a main loop and CJJ loop subjected to external flux biases $\Phi_x^q$ and $\Phi_x^{\text{\textit{cjj}}}$, respectively.
The CJJ loop is interrupted by two nominally identical Josephson junctions connected in parallel with total capacitance $C^q$ and critical current $I_c^q$. The CJJ and main loop possess inductances $L^{\text{cjj}}$ and $L^q$, respectively.  If $L^{\text{cjj}}\ll L^q$ then the RF-SQUID Hamiltonian can be written as
\begin{equation}
\label{eqn:Hrfs}
{\cal H}_{\text{rf}}(\Phi^q,\Pi^q)=\frac{1}{2C^q}(\Pi^q)^2+U(\Phi^q) \; ,
\end{equation}
\vspace{-10pt}
\begin{displaymath}
U(\Phi^q)=\frac{\left(\Phi^q-\Phi_x^q\right)^2}{2L^q}-E_J\cos\!\left[\frac{\pi\Phi_x^{\text{\textit{cjj}}}}{\Phi_0}\right]\cos\!\left[\frac{2\pi\Phi^q}{\Phi_0}\right] \; ,
\end{displaymath}

\noindent where $\Phi^q$ represents the total flux threading the main loop, $\Pi^q$ is the conjugate momentum, $E_J\equiv\Phi_0 I_c^q/2\pi$ and $\Phi _{0}=h/2e$.
This device can be operated as a qubit for $\Phi_{x}^{\text{\textit{cjj}}}\in \lbrack 0.5,1\rbrack\Phi _{0}$ and $\Phi _{x}^{\text{\textit{q}}}\approx 0$.  Denoting the ground and first excited state of ${\cal H}_{\text{rf}}$ at $\Phi_x^q=0$ by $\left|g\right>$ and $\left|e\right>$, respectively,  the qubit states can be expressed as $\left|0\right>=\left(\left|g\right>+\left|e\right>\right)/\sqrt{2}$ and $\left|1\right>=\left(\left|g\right>-\left|e\right>\right)/\sqrt{2}$.  The bias energy of Eq.~(\ref{eqn:Ham}) is given by $\epsilon =2\left\vert I_{p}^{q}\right\vert\Phi _{x}^{q}$, where the persistent current $|I_{p}^{q}|\equiv\left|\left<0\right|\Phi^q/L^q\left|0\right>\right|=\left|\left<1\right|\Phi^q/L^q\left|1\right>\right|$.  The tunneling amplitude of Eq.~(\ref{eqn:Ham}) is given by $\Delta=\left<e\right|{\cal H_{\text{rf}}}\left|e\right>-\left<g\right|{\cal H_{\text{rf}}}\left|g\right>$.  Both $\left|I_p^q\right|$ and $\Delta$ are controlled by $\Phi _{x}^{\text{\textit{cjj}}}$. Maximum $\Delta \sim\omega_p$, where $\omega_p$ is the plasma frequency of the RF-SQUID, is obtained at $\Phi _{x}^{\text{\textit{cjj}}}=\Phi_{0}/2$.  For $\Phi _{x}^{\text{\textit{cjj}}}\approx\Phi_0$ one expects $\Delta \rightarrow 0$ and the system becomes localized in $\left|0\right>$ or $\left|1\right>$ .  In this latter regime $|I_{p}^{q}|$ generates a measurable amount of flux that can be detected via an inductively coupled DC-SQUID read-out (not shown) as described in Ref.~\cite{DCSQUID,Coupler}.

The CJJ RF-SQUID from which the data presented herein were obtained
was fabricated on an oxidized Si wafer using a Nb trilayer process with wiring layers isolated by sputtered SiO$_{2}$.  The device parameters were $L^{q}=661\pm 6\,$pH, $C^{q}=146\pm 3\,$fF and $I_{c}^{q}=1.95\pm 0.05\,\mu $A.  $L^q$ was measured using a breakout structure and $C^q$ was inferred from the value of $L^q$
and measurements of the MRT peak spacing \cite{MRT}.  From the MRT spacing we also determined $\omega_p\sim10\,$GHz.  The DC-SQUID was observed to have a maximum switching current $I_{\text{sw}}^{\text{DC}}=1.9\pm0.1\,\mu$A and
the readout-qubit mutual inductance was $M_{\text{ro-q}}=16.7\pm0.2\,$ pH. The device was mounted in an Al box in a dilution refrigerator and all on-chip cross couplings were
calibrated in-situ as described in Ref.~\cite{Coupler}.


Our experimental procedure is a variant of the MRT technique first
developed by \textrm{Rouse, Han, and Lukens} \cite{MRT}.  We exploit $\Phi _{x}^{\text{\textit{cjj}}}$ to modulate $\Delta$ using a bias line with bandwidth $\sim 5\,$MHz.
A depiction of the control sequence and the evolution of the qubit potential $U(\Phi^q)$ are shown in Fig.~\ref{fig:ControlSequence}.  The state of the qubit is initialized by setting $\Phi _{x}^{\text{\textit{cjj}}}\approx \Phi_{0}/2$ (i) and then tilting $U(\Phi^q)$ via $\Phi _{x}^{\text{\textit{q}}}$ to an initial value of
$\Phi _{i}^{\text{q}}$ (ii). Slowly raising $\Phi
_{x}^{\text{\textit{cjj}}}$ to $\Phi_0$ in the presence of the tilt
traps the system in its groundstate (iii). With tunneling
suppressed, $\Phi _{x}^{\text{\textit{q}}}$ is reset to a target
value $\Phi _{f}^{\text{q}}$ (iv). Thereafter $\Phi
_{x}^{\text{\text{\textit{cjj}}}}$ is lowered (v) to $\Phi_{f}^{\text{\textit{cjj}}}$
for a prescribed amount of time $t$ during which the
system can tunnel from the initial state to the lowest lying
state in the opposite well (vi). Raising $\Phi_{x}^{\text{\textit{cjj}}}$
(vii) to $\Phi_0$ then localizes the qubit state in $\left|0\right>$ or $\left|1\right>$ (viii) which can then be distinguished by a single shot
readout. The probability of tunneling from the
initial to final state is then measured as a function of $t$ and
$\Phi _{f}^{q}$. For a given $\Phi_f^q$ the probability of
the system being found in $\left|0\right>$ can be calculated from
balancing $\Gamma _{01}$ and $\Gamma _{10}$: $dP_{0}/dt=-\Gamma
_{01}P_{0}(t)+\Gamma _{10}P_{1}(t)$, where $P_{0}(t)+P_{1}(t)=1$. In
the limit $t\rightarrow 0$ the system starts in a definite state and
this expression reduces to $dP_{0}/dt=-\Gamma _{01}$ ($P_{0}(0)=1$) or $\Gamma _{10}$ ($P_{1}(0)=1$).



The results shown herein were generated using a value of $\Phi_f^{\text{\textit{cjj}}}$ for which we measured $\left|I_{p}^{q}\right|=0.56\pm 0.02\,\mu $A. This particular target CJJ bias was chosen as $1/\Gamma$ varies by
nearly four orders of magnitude ($10\,\mu $s$\rightarrow 100\,$ms)
as a function of $\Phi _{f}^{\text{\textit{q}}}$ in the vicinity of
$\Phi _f^q=0$, which then takes full advantage of the dynamic range
of our apparatus.  Using the calibrated device parameters and Eq.~(\ref{eqn:Hrfs}) it was determined that the above mentioned value of $\left|I_{p}^{q}\right|$ could be achieved for $\Phi_f^{\text{\textit{cjj}}}=0.606\pm0.001\,\Phi_0$.  The tunnel splitting was then estimated to be $\Delta_0=0.10^{+0.28}_{-0.07}\,$mK.  The asymmetric error bars on
$\Delta_0$ are a consequence of its exponential sensitivity to errors in $L^q$, $C^q$, and $I_c^q$.

\begin{figure}[tbp]
\includegraphics[width=3in]{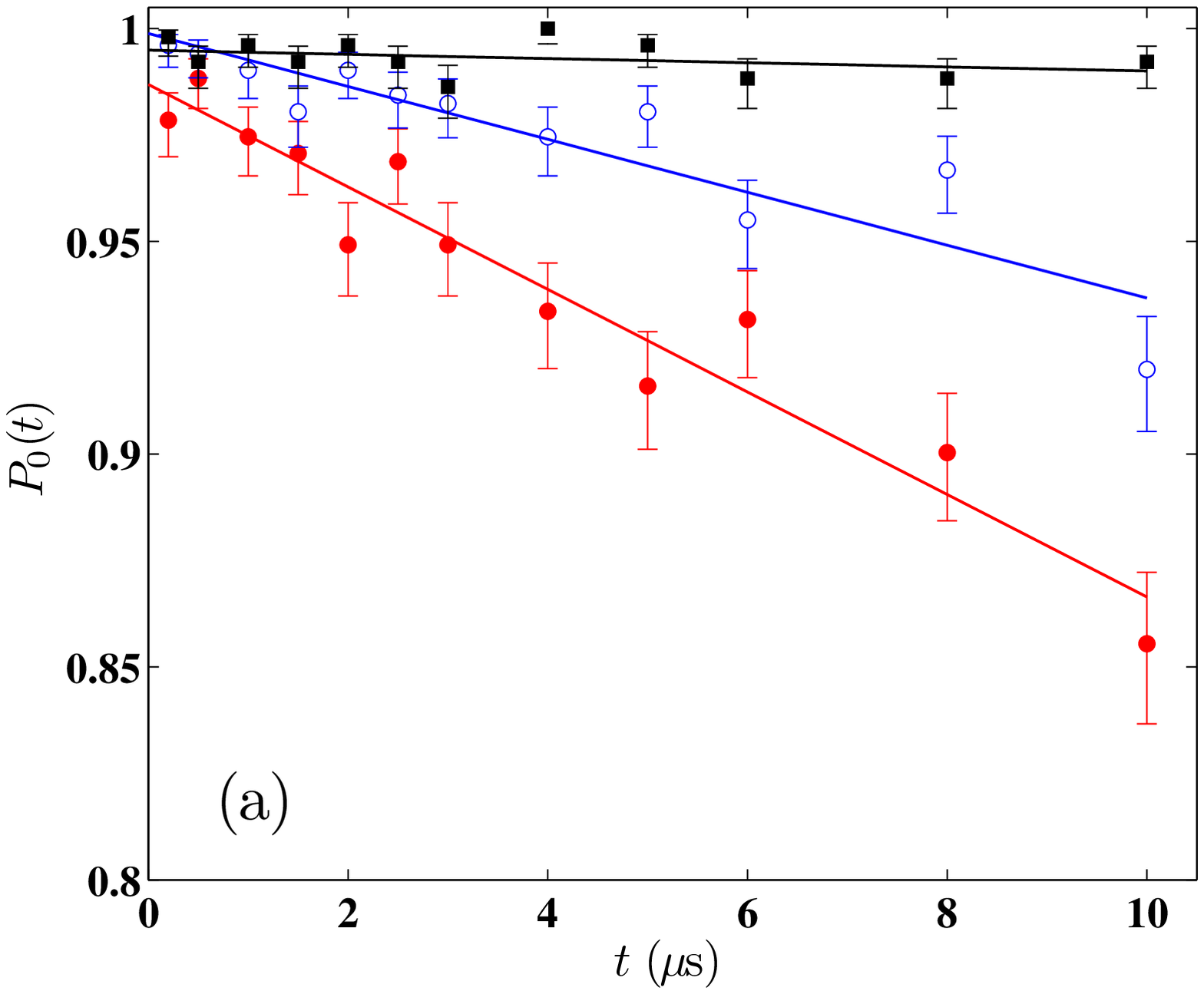}\\
\includegraphics[width=3in]{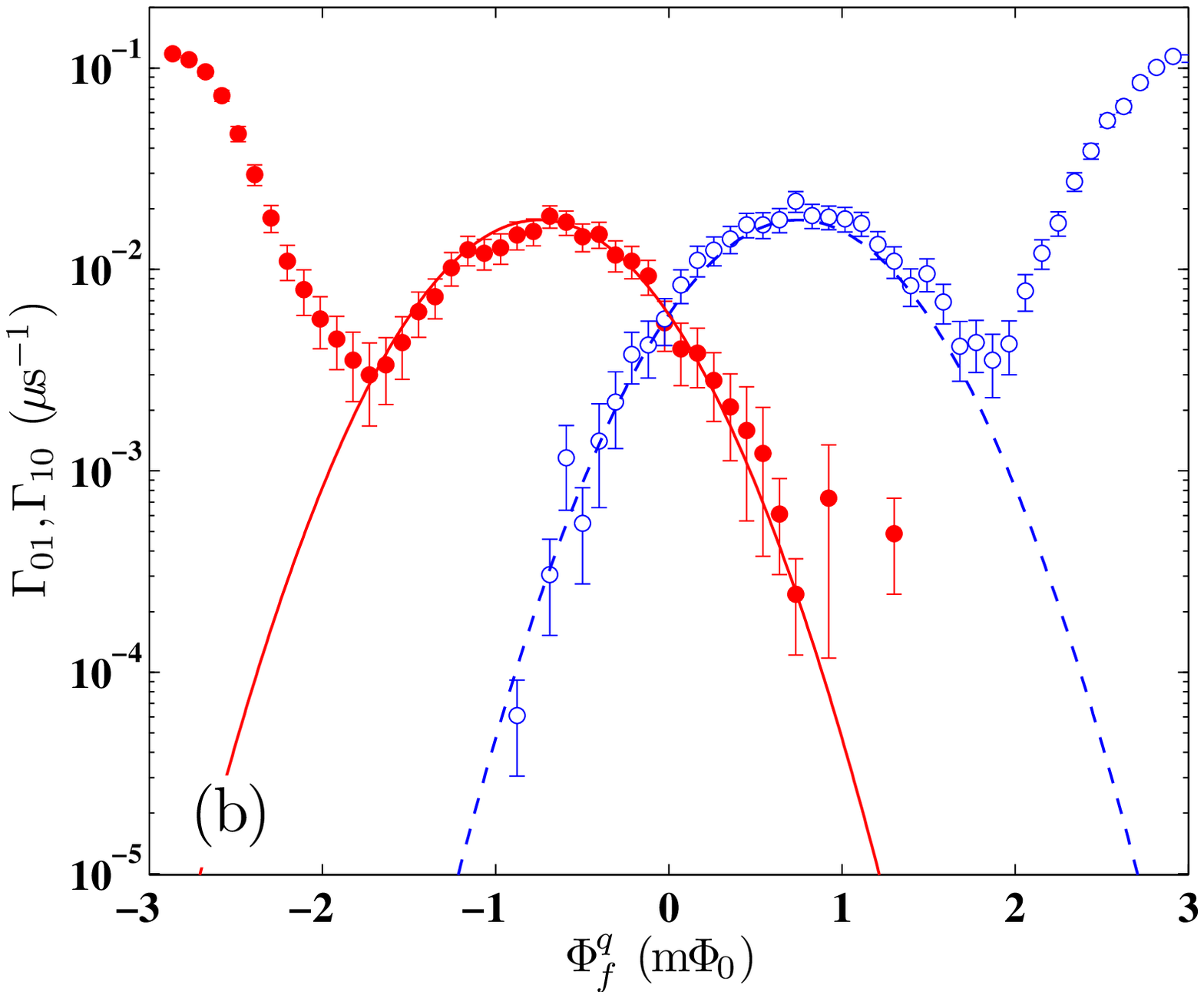}
\caption{(Color online) (a) MRT from the initial state $\left|0\right>$ versus $t$ at $T_{\text{th}}=28\,$mK.  Results are shown for $\Phi_f^q=-0.488\,$m$\Phi_0$ (solid squares), $-0.014\,$m$\Phi_0$ (hollow circles) and $0.554\,$m$\Phi_0$ (solid
circles).  Slopes of the linear fits yield
$-\Gamma_{01}(\Phi_f^q)$. (b) $\Gamma_{01}$
(hollow) and $\Gamma_{10}$ (solid) versus $\Phi_f^q$ at
$T_{\text{th}}=28\,$mK. The peaks nearest $\Phi_f^q=0$ have been fit
to Eq.~(\protect\ref{eqn:Amin1}).} \label{fig:DecayRate}
\end{figure}

We have measured $P_{0}(\Phi
_{x}^{\text{\textit{q}}},t)$ for $\Phi _{f}^{\text{\textit{q}} }\in
\left[-3,3\right]\,\text{m}\Phi _{0}$ and for the qubit initialized
with $P_{0}(0)=1$ and $P_{1}(0)=1$.
Example decay data are shown in Fig.~\ref{fig:DecayRate}(a) for $P_0(0)=1$ at
three nominal target values of $\Phi_f^{\text{q}}$ and the device
thermalized at $T_{\text{th}}=28\,$mK. A summary of initial decay rates as a
function of $\Phi_f^{\text{q}}$ for both initializations is shown in Fig.~%
\ref{fig:DecayRate}(b).  We have independently calibrated $\Phi _f^q=0$
by measuring the population statistics in the limit $t%
{\rightarrow}\infty$ and observing that the center of the resultant thermal
distribution is independent of initial state. Fitting of the
distribution to $P_0(t{\rightarrow}\infty){=}\frac{1}{2}\left[%
1-\tanh\left(\epsilon/2T_{\text{th}}\right)\right]$ yields
$T_{\text{th}}$. Note that $\Gamma_{01}$ and $\Gamma_{10}$ are
mirror images across $\Phi_f^{\text{q}}=0$ but both are asymmetric
about this point. The data suggest that $\Gamma_{01}$ and
$\Gamma_{10}$ consist of broad peaks that are displaced away from
$\Phi_f^q=0$ in the direction opposite that in which the system was
initialized. These peaks can be attributed to splitting by
noise of a single MRT peak corresponding to tunneling between the two
lowest lying states of the bistable RF-SQUID. This splitting
indicates that $S_a(\omega)\neq 0$.  Beyond
$|\Phi_f^{\text{q}}|{\gtrsim}2\,$m$\Phi_0$, the decay rates increase
due to MRT between the initialized state and the first excited state
in the opposing well.
\begin{figure}[tbp]
\includegraphics[width=3in]{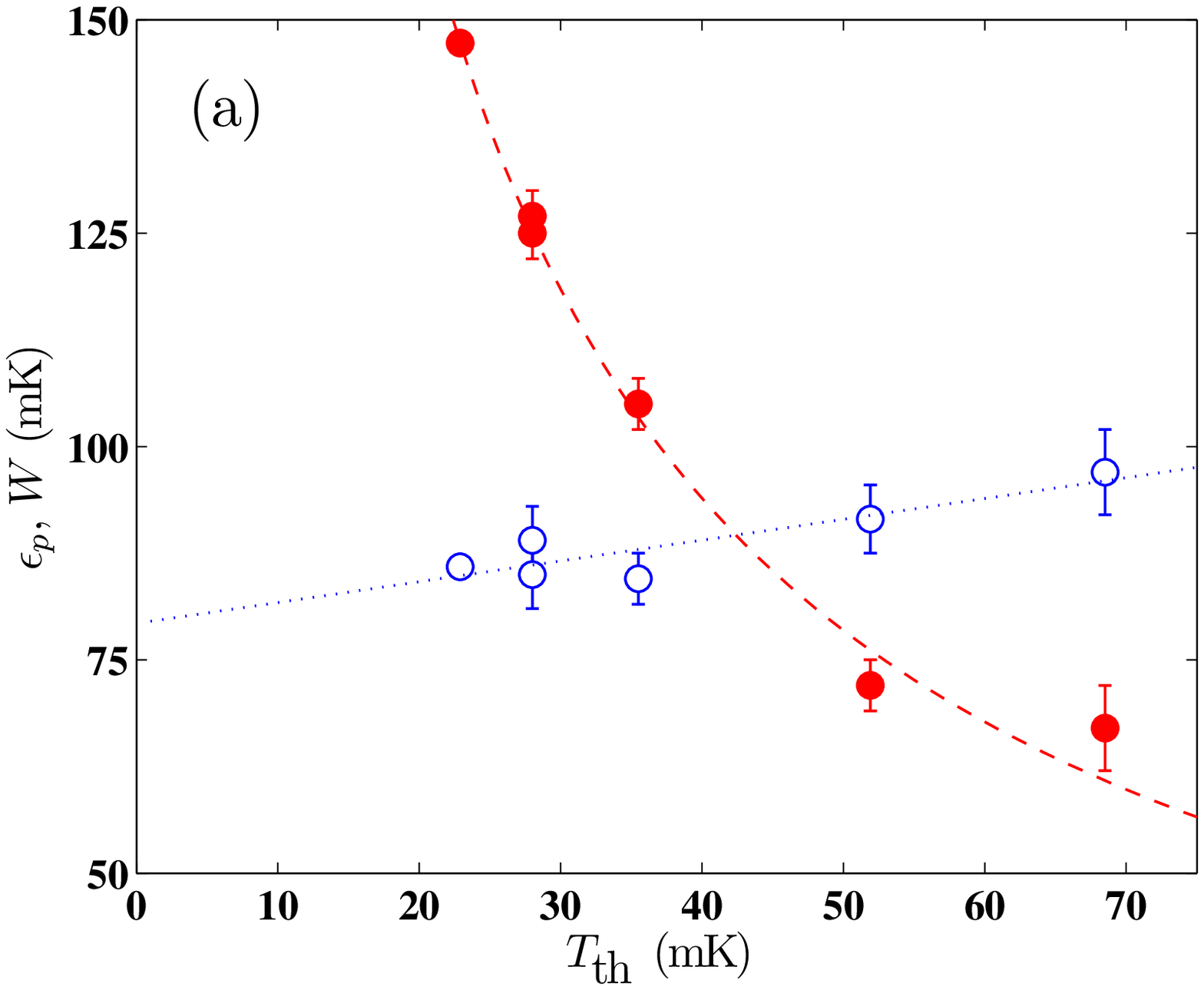}\newline
\includegraphics[width=3in]{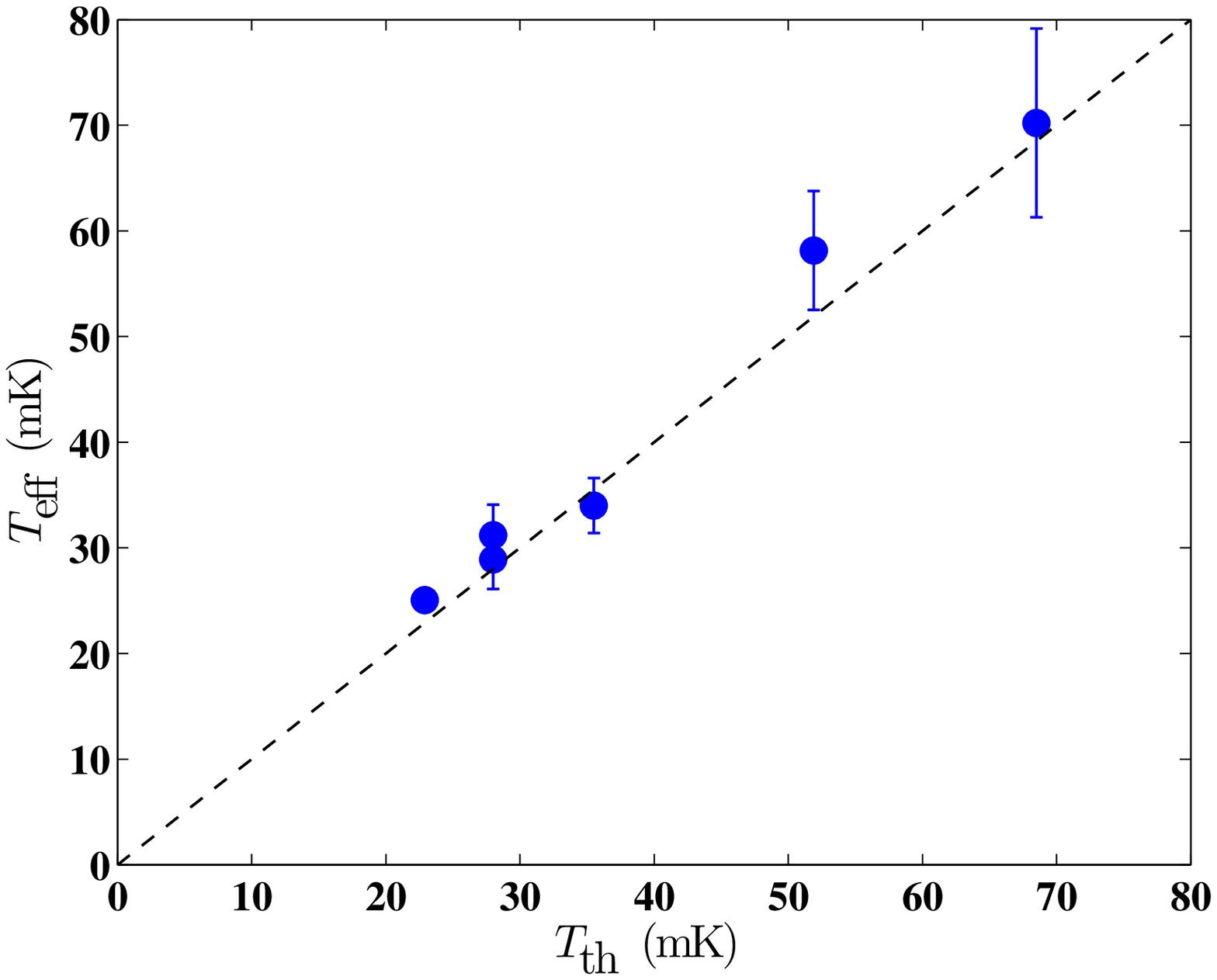}
\caption{(Color online) (a) Temperature dependence of
$\protect\epsilon_p$ (solid) and $W$ (hollow).  $W$ has been fit to a
line (dotted) and $\epsilon_p$ to a power law
$C/T_{\text{th}}^{\alpha}$ (dashed).  (b) $T_{\text{eff}}=W^2/2\epsilon_p$ versus $T_{\text{th}}$.  Dashed line indicates
$T_{\text{eff}}=T_{\text{th}}$.} \label{fig:EpAndW}
\end{figure}

The data in Fig.~\ref{fig:DecayRate}(b) have been fit to
Eq.~(\ref{eqn:Amin1}) using $W$, $\epsilon _{p}$ and $\Delta $ as
free parameters. The fact that the tunneling rate can be fit to a
Gaussian lineshape indicates that the noise is
dominated by low frequency components. We repeated the measurements
for five different temperatures and the resultant fit
values of $\epsilon _{p}$ and $ W $ are summarized in
Fig.~\ref{fig:EpAndW}(a). Here it can be seen that $W$ is only
weakly dependent upon $T$. Fitting these results to a line indicates
that $W(T\rightarrow0)\approx80\,$mK. This conclusion corroborates
the observations presented in Ref.~\cite{Lisenfeld} from phase
qubits in which the Rabi decay time is observed to have a weak
$T$ dependence as $T\rightarrow 0$.  In contrast, $\epsilon _{p}$
varies strongly with $T$, behaving as $\sim
1/T_{\text{th}}^{0.80\pm0.05}\approx 1/T$.  Finally, an explicit demonstration of the self-consistency of our analysis is shown in Fig.~\ref{fig:EpAndW}(b) where $T_{\text{eff}}$ obtained from Eq.~(\ref{Teff}) is plotted versus $T_{\text{th}}$.  Agreement between $T_{\text{eff}}$ and $T_{\text{th}}$ demonstrates the equilibrium nature of the low
frequency flux noise in our system.


\begin{figure}[tbp]
\includegraphics[width=3in]{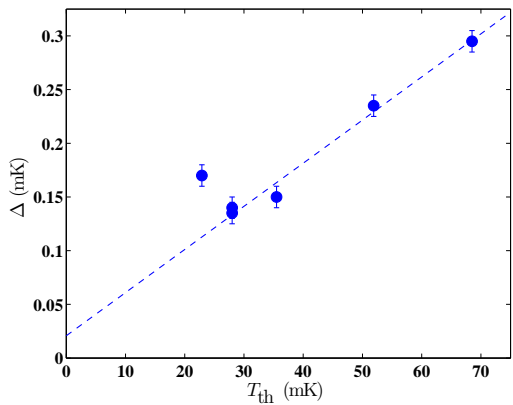}
\caption{Temperature dependence of $\Delta$. Dashed line is a fit.} \label{fig:Delta}
\end{figure}

Figure \ref{fig:Delta} shows the variation of $\Delta$ with
temperature.  Note that the results are comparable to $\Delta_0$ as estimated from qubit parameters, but that the uncertainty in $\Delta_0$ makes it difficult to draw any quantitative conclusions from such a comparison.  Nonetheless, the data suggest that $\Delta\propto T_{\text{th}}$.  It is known that high frequency flux noise can lead to renormalization of $\Delta$ \cite{weiss} and that a polynomial dependence of $\Delta$ on $T$ results if the environment possesses an ohmic density of states at high frequency \cite{Leggett}. A similar combination of sharply peaked low frequency noise, as required for Eq.~(\ref{eqn:Amin1}), and ohmic high frequency noise has been reported for other qubits \cite{1overF,Astafiev}.

In addition to studying MRT between the two lowest
lying states, we have measured MRT from the lowest state in the
initial well to higher levels in the final well.  It was
observed that the tunneling rate versus $\Phi_f^q$ could be fit to a
lineshape whose width increased monotonically with the
number of levels below the target level, as anticipated in
Refs.~\cite{Averin00} and \cite{Amin1}. The width of the peak for
tunneling to the 50-th level in the final well was roughly twice that for the lowest order MRT peak. This weak dependence on the number of states below the target level indicates the relative weakness of $S(\omega)$ at high frequency.

{\it Conclusions:} Measurements of MRT between the two lowest energy states of a bistable RF-SQUID have been used to characterize flux noise.  Analysis indicates that the noise source is a quantum mechanical environment in thermal equilibrium whose spectral density is sharply peaked at low frequency.

We thank J.~Hilton, G.~Rose, C.J.S.~Truncik, A.~Tcaciuc and
F.~Cioata for useful discussions. Samples were fabricated by the
Microelectronics Laboratory of the Jet Propulsion Laboratory,
operated by the California Institute of Technology under a contract
with NASA.



\begin{thebibliography}{99}
\bibitem{Noise} Typical amplitudes of $1/f$ flux noise
$\delta \Phi \sim 10^{-4} \Phi_0 $ (consistent with the results herein)
and charge noise $\delta Q \sim 10^{-4} e $ define an
impedance $\delta \Phi/ \delta Q \sim 10$ k$\Omega$.
Since the impedance of our flux qubit $\sqrt{L/C}\sim $ 100 $\Omega$
is much smaller than the above value, flux noise provides a stronger
decoherence mechanism than charge noise.
\end{thebibliography}
\end{document}